\documentclass[conference,9pt]{IEEEtran}

\usepackage{amsmath,amssymb,amsfonts}
\usepackage{graphicx}
\usepackage{textcomp}
\usepackage{amsmath}
\usepackage{multirow}
\usepackage{hyperref}       
\usepackage{url}            
\usepackage{booktabs}       
\usepackage{amsfonts}       
\usepackage{nicefrac}       
\usepackage{microtype}      
\usepackage{xcolor}         
\usepackage{algorithm}
\usepackage{algpseudocode}
\usepackage{amsmath}
\usepackage{cite}
\usepackage{tikz}

\usepackage{orcidlink}
\usepackage{comment}
\usepackage{amssymb}
\usepackage{braket}
\usepackage{cite}
\usepackage{xcolor}
\def\BibTeX{{\rm B\kern-.05em{\sc i\kern-.025em b}\kern-.08em
    T\kern-.1667em\lower.7ex\hbox{E}\kern-.125emX}}

\usepackage[compact]{titlesec}

\titlespacing\section{0pt}{0.3\baselineskip}{0.2\baselineskip}
\titlespacing\subsection{0pt}{0.2\baselineskip}{0.1\baselineskip}
\titlespacing\subsubsection{0pt}{0.15\baselineskip}{0.1\baselineskip}

\begin{document}
\pagestyle{plain}

\title{RobQFL: Robust Quantum Federated Learning in Adversarial Environment}

 \author{\IEEEauthorblockN{Walid El Maouaki\textsuperscript{1,2}, Nouhaila Innan\textsuperscript{2,3}, Alberto Marchisio\textsuperscript{2,3}, Taoufik Said\textsuperscript{1}, \\ Muhammad Shafique\textsuperscript{2,3}, and Mohamed Bennai\textsuperscript{1}}
\IEEEauthorblockA{\textsuperscript{1}Quantum Physics and Spintronic Team, LPMC, Faculty of Sciences Ben M'sick,\\ Hassan II University of Casablanca,
Morocco\\
\textsuperscript{2}eBRAIN Lab, Division of Engineering, New York University Abu Dhabi (NYUAD), Abu Dhabi, UAE\\
\textsuperscript{3}Center for Quantum and Topological Systems (CQTS), NYUAD Research Institute, NYUAD, Abu Dhabi, UAE\\ walid.elmaouaki-etu@etu.univh2c.ma, nouhaila.innan@nyu.edu, alberto.marchisio@nyu.edu, taoufik.said@univh2c.ma, \\muhammad.shafique@nyu.edu, mohamed.bennai@univh2c.ma\\
 }
 }
\maketitle

\begin{abstract}

Quantum Federated Learning (QFL) merges privacy-preserving federation with quantum computing gains, yet its resilience to adversarial noise is unknown. We first show that QFL is as fragile as centralized quantum learning. We propose Robust Quantum Federated Learning (RobQFL), embedding adversarial training directly into the federated loop. RobQFL exposes tunable axes: client coverage $\gamma$ (0-100\%), perturbation scheduling (fixed-$\varepsilon$ vs $\varepsilon$-mixes), and optimization (fine-tune vs scratch), and distils the resulting $\gamma \times \varepsilon$ surface into two metrics: Accuracy–Robustness Area and Robustness Volume.
On 15-client simulations with MNIST and Fashion-MNIST, IID and Non-IID conditions, training only 20--50\% clients adversarially boosts $\varepsilon \leq 0.1$ accuracy $\sim$15 pp at $< 2$ pp clean-accuracy cost; fine-tuning adds 3--5 pp. With $\geq$75\% coverage, a moderate $\varepsilon$-mix is optimal, while high-$\varepsilon$ schedules help only at 100\% coverage. Label-sorted non-IID splits halve robustness, underscoring data heterogeneity as a dominant risk.

\end{abstract}

\begin{IEEEkeywords}
Quantum Machine Learning, Quantum Federated Learning, Adversarial Robustness, Quantum Neural Networks, Non-IID Data
\end{IEEEkeywords}


\section{Introduction}
Federated Learning (FL) is a distributed optimization protocol in which a population of clients trains a common model by transmitting parameter updates, rather than raw data, to a central server \cite{zhang2021survey}. This design mitigates legal and ethical risks associated with data centralization, reduces communication overhead, and enables on-device personalization.
QML investigates algorithms that exploit quantum resources—superposition, entanglement, and interference—to construct learning models, such as Quantum Neural Networks (QNNs) and quantum kernel methods \cite{biamonte2017quantum, cerezo2022challenges,abbas2021power,innan2023enhancing, zaman2024comparative, zaman2024studying, innan2024variational,innan2025next}. Although current quantum hardware is noisy and small-scale, theoretical and empirical studies suggest that certain tasks may admit polynomial- or exponential-scale advantages in expressivity or sample efficiency when executed on quantum processors. Proof-of-concepts already span diverse areas—from cancer diagnosis~\cite{maouaki2024quantum,khan2024brain} and drug discovery~\cite{blunt2022perspective} to finance~\cite{orus2019quantum,innan2024financial1,innan2024financial,alami2024comparative, zaman2024po, innan2024lep, choudhary2025hqnn,innan2025optimizing, innan2025qnn,dave2025sentiqnf,innan2025quantum}, and cybersecurity~\cite{el2024quantum}—showing QML’s broad practical reach.
Quantum Federated Learning (QFL) merges the privacy-preserving structure of FL with the potential speed-ups of QML, which may one day solve some tasks faster than classical models, and federated learning, which keeps each client’s data on-device \cite{sawaika2025privacy,innan2024qfnn}. But stronger privacy does not mean stronger security. When an attacker adds small, well-chosen noise to the input, both classical and quantum models can fail badly. Recent studies confirm that QNNs suffer from adversarial susceptibility just like their classical counterparts \cite{el2024robqunns,el2024advqunn,el2025designing,lu2020quantum}. Our analysis reveals that QFL models are as vulnerable as centralized quantum models, collapsing under mild perturbations regardless of the training paradigm. Yet today, there is limited defense designed for the federated quantum setting. Classical adversarial training cannot be copied over as is, as data representation differs, and gradient-based perturbation is prohibitively expensive.

To bridge these gaps, we build upon the recently proposed Quantum Federated Adversarial Learning (QFAL) \cite{maouaki2025qfal} by introducing novel enhancements that further strengthen collaborative adversarial defenses within QFL workflows.
The proposed RobQFL framework introduces three targeted enhancements to the standard QFL pipeline: (i) coverage parameter ($\gamma$), enabling adversarial training to be selectively applied to a specified subset of clients; (ii) a flexible $\varepsilon$-scheduler, accommodating both fixed $\varepsilon$ and cyclic multi-$\varepsilon$ perturbation strategies to reflect varying adversarial threat intensities; and (iii) dual optimization strategies, allowing either fine-tuning of a pre-trained clean model or training a robust model from the ground up. To facilitate standardized evaluation and comparison across defense strategies, the resulting $\gamma \times \varepsilon$ accuracy landscape is further summarized using two aggregate metrics: the Accuracy-Robustness Area (ARA) and the Robustness Volume (RV).

\subsection*{Contributions}
\begin{itemize}
    \item We give the first head-to-head test of centralized vs. federated quantum learning under adversarial attacks and show that simple federation brings no extra robustness.

\item We design and implement an adversarial-training protocol tailored to QFL, with flexible client coverage and perturbation schedules.

\item We introduce two novel metrics, ARA and RV, to comprehensively evaluate the effectiveness of defenses across diverse attack intensities and coverage levels. These metrics provide a clearer picture of defense reliability under varying adversarial scenarios.

\item Experiments on 15 simulated quantum clients reveal that defending even $20-50\%$ of clients increases accuracy under moderate attacks by $\sim$15 pp while cutting clean accuracy by $<2$ pp, and that data heterogeneity is now the main barrier to secure QFL.

\end{itemize}

\section{Background and Related Work}

\subsection{Machine Learning Security}

Machine learning (ML), especially deep learning models, is known to be highly susceptible to adversarial attacks, where carefully crafted perturbations to input data lead to incorrect predictions while remaining virtually indistinguishable to humans~\cite{szegedy2013intriguing}. These vulnerabilities are particularly critical in domains such as healthcare, finance, and autonomous driving, where erroneous model outputs can have severe consequences.

One of the most widely adopted attack strategies is the Projected Gradient Descent (PGD) attack~\cite{madry2017towards}. PGD operates by iteratively adjusting input samples to maximize model loss, while constraining the perturbations within a bounded region to maintain input validity. Formally, for a classifier $f_\theta$ with parameters $\theta$, a clean input $x$, and its label $y$, PGD generates adversarial examples $x'$ that maximize the classification error while satisfying a norm-bound constraint on perturbation magnitude. The iterative nature of PGD allows for increasingly effective perturbations that can consistently fool even robust models.

To mitigate such threats, adversarial training has emerged as one of the most effective defense mechanisms~\cite{madry2017towards}. This approach incorporates adversarially perturbed data into the training process, allowing models to learn more robust representations. Although adversarial training improves model resilience, it comes at the cost of higher computational demands and often leads to slight decreases in performance on unperturbed data~\cite{tsipras2018robustness}. 

In the context of QFL, integrating adversarial training presents additional complexity due to inherent quantum constraints such as decoherence, shot noise, and limited qubit counts~\cite{maouaki2025qfal}. Our work builds on these concepts by adapting adversarial training strategies to the federated quantum setting, aiming to bolster model robustness against adversarial manipulations in distributed quantum environments.

\subsection{Quantum Federated Learning}

QML merges quantum computing with classical ML frameworks to address complex computational challenges and potentially achieve superior efficiency~\cite{schuld2015introduction, zaman2023survey, kashif2024computational}. A central component of QML is the QNNs, which utilize parameterized quantum circuits (PQCs) as the core learning architecture. In QNNs, quantum gates are controlled by trainable parameters, forming layered quantum circuits that modify quantum states. After applying these transformations, quantum measurements are performed to obtain predictions.

A crucial step in QNNs is encoding classical data into quantum states. Common encoding schemes include basis encoding, angle encoding, and amplitude encoding, which map classical data vectors into quantum states. QNNs are typically trained in supervised learning setups by minimizing loss functions such as the Mean Squared Error (MSE).

Since quantum measurements collapse the state, multiple measurements (shots) are required to estimate these expectation values accurately, introducing statistical noise into gradient estimation. Training QNNs typically relies on classical optimization techniques. The parameter-shift rule enables efficient gradient computation for specific parameterized gates.

Extending these principles, Quantum Federated Learning (QFL) adapts the federated learning paradigm to quantum models. In QFL, multiple quantum clients collaboratively train a global model without sharing raw quantum data, thus preserving privacy while addressing constraints like the no-cloning theorem. Each client $k$ maintains a local QNN with parameters $\theta_k^{(t)}$ and contributes updates to the global model via aggregation:
$$
    \theta^{(t+1)} = \sum_{k=1}^{K} w_k \theta_k^{(t)},
$$
where the weights $w_k$ typically reflect the relative data sizes or computational resources of each client, normalized so that $\sum_{k=1}^{K} w_k = 1$. The global model is then redistributed, and this process continues until convergence.

Initial work in QFL validated the feasibility of federated learning with hybrid quantum-classical models, showing faster convergence rates with comparable accuracy~\cite{chen2021federated}. Subsequent research explored QFL with variational quantum circuits while maintaining privacy~\cite{huang2022quantum}, and federated quantum neural networks were applied to healthcare and genomics with promising results~\cite{innanqfl}. Extensions such as FL-QDSNNs introduced spiking neural networks into QFL to enhance scalability and privacy in distributed quantum systems~\cite{innanqsnn}.

Security in QFL has also been investigated from multiple angles. Techniques like blind quantum computation~\cite{li2021quantum}, quantum fuzzy federated learning for non-IID data~\cite{qu2024quantum}, quantum differential privacy~\cite{rofougaran2024federated}, and homomorphic encryption~\cite{dutta2024mqfl,dutta2024federated} have all been proposed to strengthen privacy guarantees. Communication-efficient methods such as federated quantum natural gradient descent have further optimized training dynamics~\cite{qi2024federated}.

While privacy has received considerable attention, adversarial robustness in QFL remains underexplored. Federated models in sensitive applications such as autonomous vehicles and cybersecurity face threats from poisoning attacks and Byzantine failures. Solutions like quantum-behaved particle swarm optimization for vehicular networks~\cite{qi2023optimizing} and quantum-inspired federated averaging for cyber-attack detection~\cite{subramanian2024hybrid} have demonstrated partial resilience. Moreover, Byzantine fault tolerance in QFL has been studied through adaptations of classical approaches~\cite{xia2021defending}. However, the vulnerability of QFL models to evasion attacks that exploit quantum phenomena such as superposition, entanglement, and measurement uncertainty is still largely unexplored.

Unlike classical adversarial techniques such as FGSM~\cite{goodfellow2014explaining} and PGD~\cite{madry2017towards}, their applicability to QFL remains unclear. Motivated by these gaps, our work aims to conduct a comprehensive analysis of adversarial attacks in QFL and investigate their robustness to adversarial attacks in quantum federated systems.

\subsection{Related Work}
Federated learning (FL) is known to be vulnerable to adversarial manipulation of data and model updates. Studies show that FL models can be as fragile as centrally trained models when facing adversarial examples. To address such threats, classical FL research has explored robust aggregation methods and adversarial training, but these defenses often involve trade-offs that hurt model accuracy, especially under non-IID data distributions \cite{zhang2023delving}.

Quantum Federated Learning (QFL) aims to combine these ideas with quantum neural networks (QNNs). Most work so far tackles privacy rather than robustness. Li et al. hide gradients with quantum states, showing that blind inner-product estimation thwarts gradient-inversion attacks while keeping bandwidth low \cite{li2024privacy}. Li and Deng propose quantum homomorphic encryption so that a server can train on encrypted qubits, gaining formal privacy guarantees and lower client overhead \cite{li2025quantum}. Chen et al. inject Laplace noise into parameter updates and demonstrate that moderate noise improves QFL accuracy under simulated quantum channel errors and adversarial perturbations \cite{chen2024robust}.
Hybrid schemes that mix encryption with quantum resources are also emerging. The MQFL-FHE framework couples fully homomorphic encryption with QNNs to recover performance lost by classical FHE during aggregation \cite{dutta2024mqfl}.

Although these studies enhance privacy or baseline robustness, none systematically analyze adversarial training coverage, perturbation scheduling, or non-IID effects in QFL. In particular, prior work leaves open three critical practical questions: How much adversarial coverage $(\gamma)$ is actually needed? Which perturbation radii $(\varepsilon)$ should be used, and should they be uniform or client-specific? How do these choices interact with quantum-specific cost constraints? Bridging this gap requires adapting classical robustness insights to quantum circuits while respecting these constraints. Furthermore, while client-side adversarial training improves resilience, it often raises communication costs and lowers clean-data accuracy. RobQFL framework addresses these gaps by (i) introducing partial-coverage adversarial training to cap communication and circuit-execution overhead, (ii) comparing fixed versus mixed $\varepsilon$-schedules that reflect heterogeneous client resources, and (iii) evaluating all schemes under realistic non-IID data splits. This yields the first quantitative design rules for balancing clean-data performance, robustness, and quantum-hardware cost in federated settings.

Previous work \cite{maouaki2025qfal} was the first to integrate adversarial training into the quantum federated learning (QFL) framework. However, it reports results on only a single IID dataset (MNIST), keeps the same $\epsilon$ in every round, and does not study how non-IID splits, dynamic $\epsilon$ schedules, or alternative optimization regimes affect robustness. These gaps leave open questions about real-world deployment. Addressing them calls for experiments with partial-coverage adversarial training, variable $\epsilon$ curricula, diverse data distributions, and richer optimization strategies—issues the present work tackles to move QFL robustness research from theory toward practical security.

\section{Methodology and Experimental Setup}
\label{sec:methodology}

Let $\mathcal{K}=\{1,\dots,K\}$ be the set of $K=15$ quantum clients, each owning a private classical dataset $\mathcal{D}_k$.  
A single parameterized quantum circuit (PQC) $f(\mathbf{x};\boldsymbol{\theta})$ acts as the hypothesis class.  
The circuit embeds an input vector $\mathbf{x}\in\mathbb{R}^d$ into the amplitudes of a 4-qubit register, applies two layers of a hardware-efficient ansatz $U(\boldsymbol{\theta})$, and returns an expectation value that is mapped to class logits.  
Training proceeds in the Quantum-Federated Learning (QFL) setting: every communication round $t$, each selected client performs local gradient descent on its loss $\mathcal{L}_k$ and transmits updated parameters $\boldsymbol{\theta}_k^{(t)}$ to the server, which aggregates them through Federated Averaging:
$$
\boldsymbol{\theta}^{(t+1)}
\;=\;
\frac{1}{\lvert S_t\rvert}
\sum_{k\in S_t}
\boldsymbol{\theta}_k^{(t)}.
$$
This protocol preserves data privacy while allowing a single quantum model to be trained collaboratively.

To address adversarial threats, we introduce an untargeted PGD adversary that perturbs each clean sample inside an $\ell_\infty$ ball of radius $\epsilon\in[0,0.5]$:
\[
x^* \;=\; x \;+\;
\arg\max_{\|\delta\|_\infty \le \epsilon}
\mathcal{L}\bigl(f(x+\delta;\theta),\,y\bigr).
\]
Adversarial training (AT) is then applied on only a fraction $\gamma \in \{0\%,20\%,50\%,75\%,100\%\}$ of the clients, thereby isolating coverage as an explicit control variable. Within every covered client, we adversarially perturb $\rho = 50\%$ of each mini-batch. We evaluate two adversarial-training schedules. In the first, which we call \emph{fixed-$\epsilon$ AT}, every client applies the same perturbation radius (either $\epsilon = 0.1$ or $\epsilon = 0.3$) throughout training. In the second, \emph{mixed-$\epsilon$ AT}, we introduce a length-$m$ vector of radii $\Xi = [\epsilon_1,\dots,\epsilon_m]$ and assign its elements to the $K$ adversarially trained clients in a cyclic fashion. Concretely, client $k$ uses

$$\epsilon_k=\Xi[(K \bmod m)-1],$$

in our experiments $m = 3$ where we explore three specific mixes: a low perturbations mix, $\Xi_1 = [0.01,\,0.02,\,0.05]$, a moderate mix, $\Xi_2 = [0.10,\,0.15,\,0.2]$, and strong mix, $\Xi_3 = [0.3,\,0.4,\,0.5]$. This cyclic assignment allows us to probe how heterogeneous threat levels across clients interact with different coverage rates in a federated learning setting.

To disentangle the effect of when adversarial training is injected, we design two optimization regimes. (i) Fine‑tune. We first obtain a reference QFL model by running 50 communication rounds on clean data; this mimics the common scenario in which a consortium has already converged on a high-performing, but non-robust, network. We then resume training for a further 30 rounds under the selected adversarial‑training (AT) schedule, using the previously learned parameters $\theta^{clean}$ as a warm‑start. (ii) Scratch. In contrast, this regime discards any warm‑start and trains the model from random initialization for 20 communication rounds entirely under the same AT schedule, reflecting deployments that seek robustness from the outset. Comparing these two trajectories allows us to quantify the trade‑offs between retrofitting robustness onto an existing model and building it in from the beginning.

To holistically evaluate adversarial robustness under varying training conditions, we propose three complementary metrics. First, the \emph{standard accuracy} \(A(\gamma,\epsilon)\) measures the model’s top-1 classification performance on 1000 test samples under adversarial perturbations of strength \(\epsilon\), given a fixed adversarial training coverage \(\gamma\).

Building on this, the \emph{Accuracy-Robustness Area} (ARA) quantifies robustness across perturbation strengths for a specific coverage level \(\gamma\), defined as
\[
\mathrm{ARA}(\gamma)
= \frac{1}{\epsilon_{\max}}
\int_{0}^{\epsilon_{\max}}
\frac{A(\gamma,\epsilon)}{100}
\,d\epsilon.
\]

This metric integrates the normalized accuracy over the perturbation range \(\epsilon\in[0,0.5]\), penalizing sharp drops in performance under stronger attacks. Dividing by $\epsilon_{\max}=0.5$ converts the raw area into the mean accuracy over the tested $\epsilon$-range, so ARA is scale-free and always lies in $[0,1]$; $1$ means the model kept $100\%$ accuracy for every $\epsilon$, $0$ means it failed completely.

To generalize across both coverage and perturbation, we introduce the \emph{Robustness Volume} (RV):
\[
\mathrm{RV}
= \frac{1}{\epsilon_{\max}\,\gamma_{\max}}
\int_{0}^{\gamma_{\max}}
\int_{0}^{\epsilon_{\max}}
\frac{A(\gamma,\epsilon)}{100}
\,d\epsilon\,d\gamma.
\]

By collapsing the whole surface \(A(\gamma,\epsilon)\) into a single number in \([0,1]\), RV lets you compare robustness across different training setups. Together, these metrics disentangle the effects of attack strength, federated participation, and model resilience, with RV giving a clear benchmark for robustness in distributed quantum learning. The factor
$
\frac{1}{\epsilon_{\max}\,\gamma_{\max}},
$
turns the surface integral into the mean accuracy across all coverages and perturbations. RV therefore provides a single, comparable robustness score in $[0,1]$, unaffected by how wide a $\epsilon$- or coverage window is chosen.

\begin{figure*}
    \centering
    \includegraphics[width=0.7\linewidth]{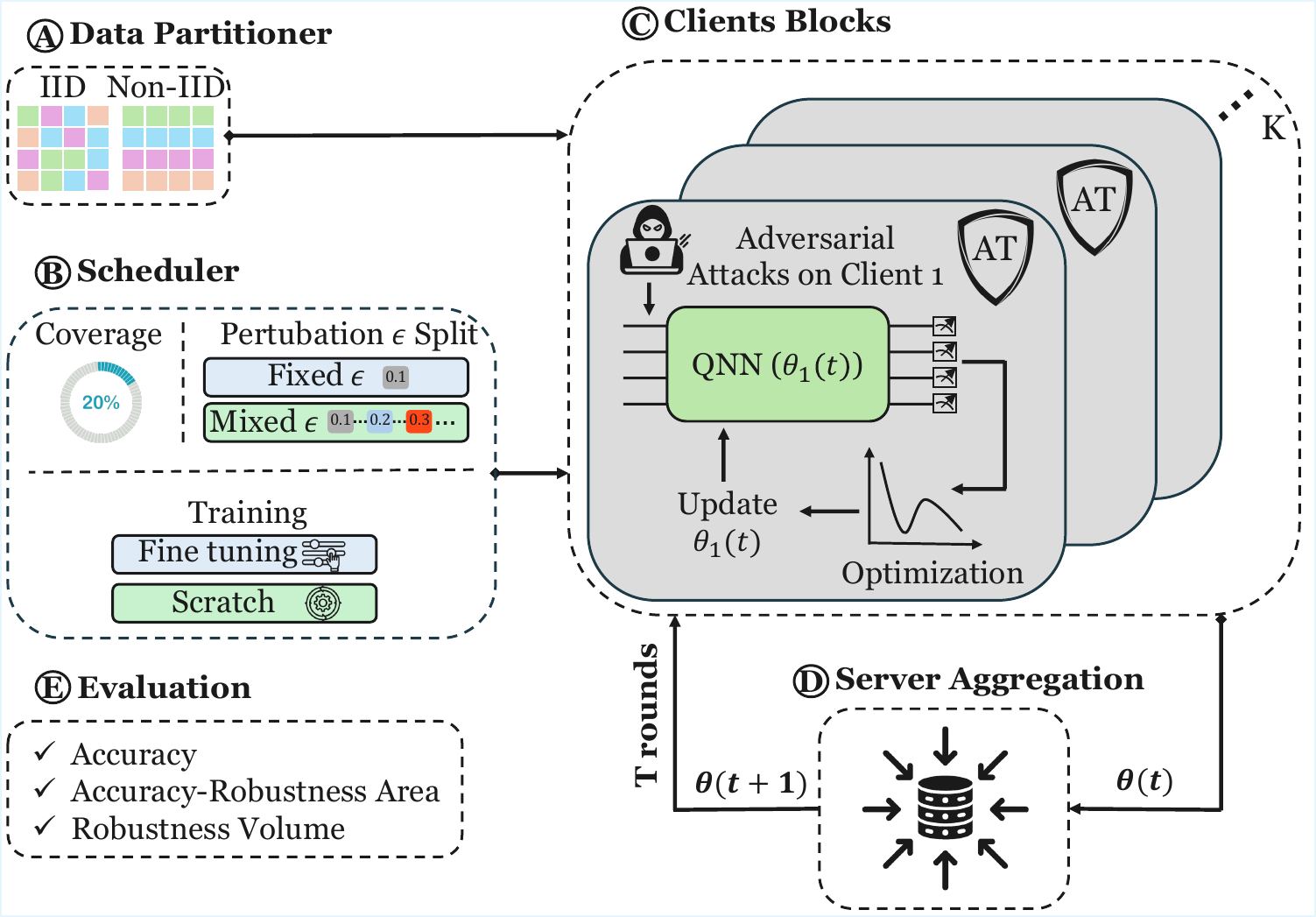}
    \caption{A) Datasets are partitioned either IID (shuffled colours) or label-sorted non-IID, then dispatched to $K$ quantum clients. B) A coverage dial selects the fraction $\gamma$ of clients that perform adversarial training (AT); each covered client follows either a \textbf{fixed-}$\varepsilon$ schedule or a \textbf{mixed-}$\varepsilon$ schedule that cycles through low, moderate, or strong perturbation radii. Two optimization regimes are supported: \textbf{fine-tune} (warm-start from a clean model) and \textbf{scratch} (random initialization). C) Inside a covered client, an N-qubit QNN $f(\cdot\,;\theta_k(t))$ is trained on minibatches where a p-fraction is perturbed by a local PGD adversary. D) After local optimization, updated parameters $\theta_k(t)$ are sent to the server and aggregated via Federated Averaging to form the global model $\theta(t+1)$. E) After the final communication round, the global model is tested on clean inputs ($\epsilon=0$) and on a sweep of adversarial perturbations. }
    \label{Methodology_QFAL}
\end{figure*}

All experiments use MNIST and Fashion‑MNIST, each reduced to 18000 training and 1000 test images. We test both an IID split (uniform shuffle) and a label‑sorted non‑IID split that yields near single‑class clients. The images are resized to dimensions of 8 by 8 pixels, and the dataset is restricted to include only the three classes {0, 1, 2}. A 4‑qubit circuit is simulated in Pennylane with parameter‑shift gradients. Refer to Fig. \ref{Methodology_QFAL} for a detailed representation of the RobQFL workflow.

By controlling coverage, perturbation schedule, and initialization while holding every other variable constant, we can attribute changes in ARA or RV directly to our proposed mixed-$\epsilon$ strategy and coverage‑aware training protocol.

\section{Results and Discussion}
\subsection{Centralized vs Decentralized}
Table \ref{tab:robustness_comparison_transposed} gives a direct baseline: a single quantum neural network trained centrally versus the same architecture deployed in our quantum-federated (QFL) setting, both evaluated under increasing adversarial perturbation~$\varepsilon$. At $\varepsilon=0$ (clean data), the two models are nearly indistinguishable (80.6\% vs.\ 79.5\%). The gap stays within 0.5--1\,pp at $\varepsilon=0.01$ and $\varepsilon=0.05$, and both curves fall sharply once $\varepsilon$ reaches $0.1$, dropping below 33\% accuracy. Beyond $\varepsilon \ge 0.2$, the collapse is complete for either training paradigm.
Decentralising the training process does not, by itself, improve robustness; if anything, QFL is as vulnerable to adversarial examples as centrally trained models. This result underscores that adversarial vulnerability is a model‑level issue, independent of where optimization takes place.
Because simple federation offers no protection, the remainder of this section focuses on targeted defenses within QFL, specifically adversarial‑training schemes with varying coverage and $\epsilon$ schedules, and shows how they shift the robustness curve back toward the clean‑accuracy regime.

\begin{table}[ht]
    \centering
    \caption{Robustness Comparison: QFL vs. Centralized across Perturbations}
    \begin{tabular}{lccccccc}
        \toprule
        \textbf{Model} & \textbf{0} & \textbf{0.01} & \textbf{0.05} & \textbf{0.1} & \textbf{0.2} & \textbf{0.3} & \textbf{0.5} \\
        \midrule
        Centralized & 80.6 & 75.9 & 58.2 & 32.9 & 0.2 & 0.0 & 0.0 \\
        QFL         & 79.5 & 75.4 & 59.6 & 24.7 & 0.0 & 0.0 & 0.0 \\
        \bottomrule
    \end{tabular}
    \label{tab:robustness_comparison_transposed}
\end{table}

\subsection{Fine tuning vs Scratch Adversarial Training}
\begin{figure}
    \centering
    \includegraphics[width=0.73\linewidth]{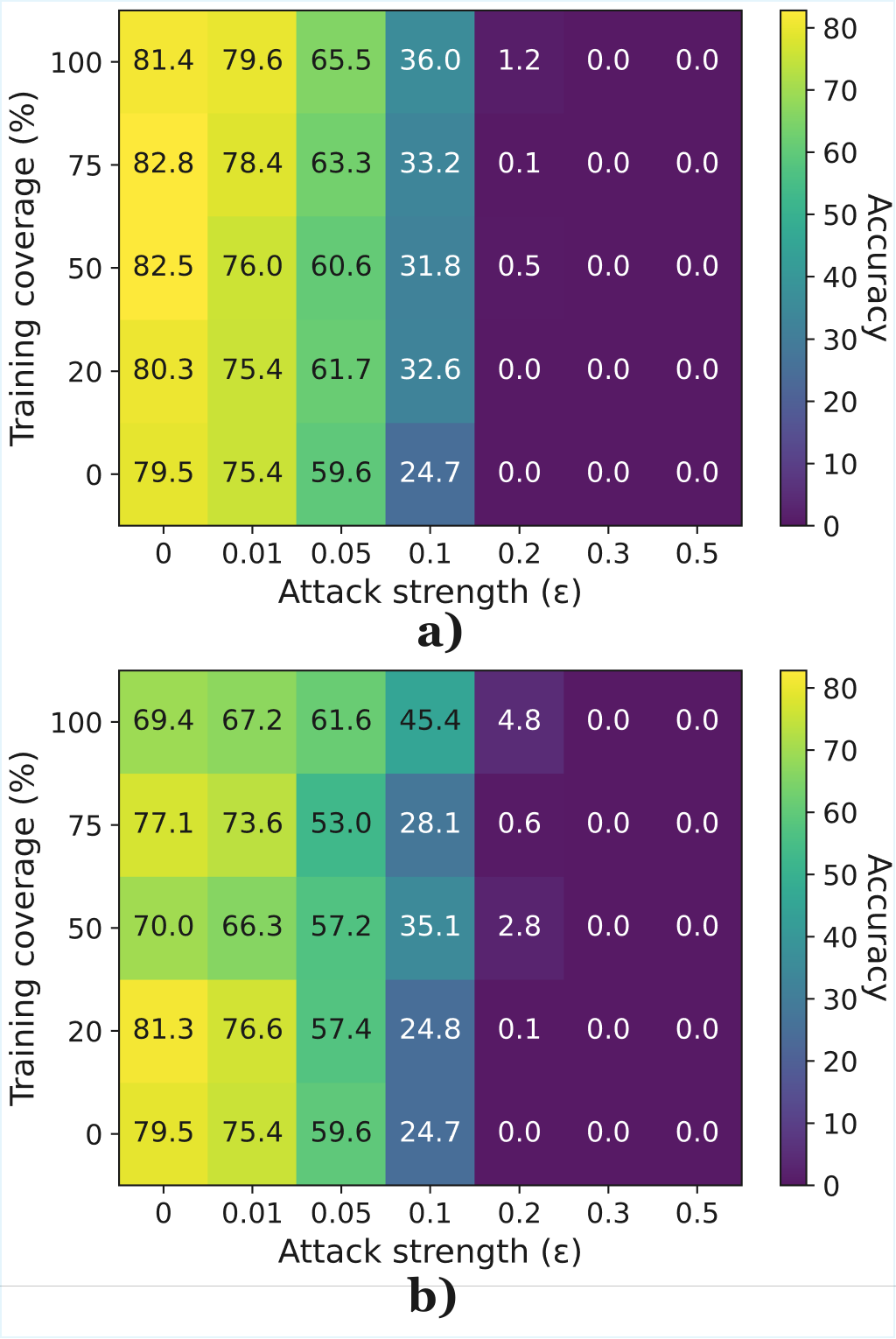}
    \caption{IID MNIST a) Fine-tuning adversarial training b) Scratch adversarial training}
    \label{fig:heatmaps_finetuning_scratch}
\end{figure}

Comparative analysis of the heatmaps, Fig \ref{fig:heatmaps_finetuning_scratch}, generated under the \emph{Fine-tune} ($\varepsilon=0.1$) and \emph{Scratch} ($\varepsilon=0.1$) experimental configurations reveals distinct trends in adversarial robustness. Both heatmaps exhibit a typical cliff effect, characterized by high model accuracy ($\varepsilon \leq 0.01$), followed by a sharp decline at $\varepsilon = 0.1$ and near-zero accuracy for $\varepsilon \geq 0.3$. Divergences, however, emerge along the training-coverage axis. The fine-tuned model, Fig \ref{fig:heatmaps_finetuning_scratch}a, demonstrates a monotonic relationship between adversarial training coverage and accuracy, with incremental improvements (evident as darker-to-lighter color gradients) as coverage increases by 25\% increments. In contrast, the scratch-trained model, Fig \ref{fig:heatmaps_finetuning_scratch}b, exhibits non-monotonic behavior: accuracy declines anomalously at 50\% coverage for weaker perturbations ($\varepsilon \leq 0.05$) before recovering at 75\% coverage, suggesting instability in optimization under partial adversarial training conditions.

Under $\varepsilon = 0$ (clean data), fine-tuning preserves baseline accuracy (79–83\% across coverages), while scratch training introduces instability, notably suffering a significant reduction in clean accuracy (69–81\%) at 100\% coverage. For $\varepsilon = 0.01$–0.05, fine-tuning consistently outperforms scratch training by margins $\geq 4$ percentage points (pp) at all coverage levels, except at 20\% coverage where scratch briefly matches fine-tune. At $\varepsilon = 0.1$, scratch training surpasses fine-tuning only under high-coverage conditions ($\geq 50\%$ of clients adversarially trained), achieving 25–45\% accuracy compared to 25–36\% for fine-tuning, though both methods show parity at low coverages ($\leq 25\%$ accuracy). For $\varepsilon \geq 0.2$, both approaches collapse ($\leq 5\%$ accuracy), with scratch marginally outperforming fine-tune under extreme attacks but remaining impractical.

Table \ref{tab:ara_volume_scratch_tuning} presents the aggregate robustness metrics which further quantify these trends. Fine-tuning achieves superior Average ARA (0.151 vs.\ 0.146) and Robustness Volume (0.151 vs.\ 0.144), reflecting $\approx$3–5\% relative gains in overall adversarial resilience.

\begin{table}[ht]
    \centering
    \caption{Robustness and mean ARA for Fine tuning and scratch adversarial training}
    \begin{tabular}{lcc}
        \toprule
Experiment & Robustness Volume & ARA\_mean \\
\midrule
Fine-tune $\epsilon=0.1$ & 0.151307 & 0.150746 \\
Scratch $\epsilon=0.1$    & 0.144324 & 0.146190 \\
\bottomrule
    \end{tabular}
    \label{tab:ara_volume_scratch_tuning}
\end{table}

The divergence in performance is attributable to training stability. Fine-tuning leverages a pre-trained model, enabling gradual adaptation to adversarial perturbations while preserving baseline accuracy. This results in smooth, monotonic improvements as adversarial training coverage expands. Conversely, scratch training’s full re-initialization on perturbed data introduces stochasticity during optimization, destabilizing clean accuracy (e.g., the 100\% coverage dip) while enabling specialized feature learning under high-$\varepsilon$ conditions with near-full coverage ($\geq 75\%$). However, fine-tuning dominates in practical settings, maintaining consistently higher accuracy ($\varepsilon \leq 0.05$, $\leq 75\%$ coverage) and reliability across heterogeneous client deployments.

For quantum federated adversarial learning on MNIST, a defence must cope with heterogeneous client participation and moderate attack strengths. Under these realistic conditions, fine-tuning provides a more reliable and generally superior solution. Scratch training provides limited gains only under stringent conditions (high adversarial coverage/predictable high-$\varepsilon$ attacks), sacrificing clean-data accuracy and exhibiting instability otherwise. As a consequence, throughout the remainder of this work, we adopt fine-tuning as our primary optimization strategy to ensure robust and stable quantum federated adversarial learning.

\subsection{Epsilon Split Adversarial Training}

We assessed five adversarial-training (AT) schedules in the same 15-client QFL setting: two single-\(\epsilon\) baselines (\(\epsilon = 0.1\) and \(\epsilon = 0.3\)) and three cyclic schedules that assign low perturbations \([0.01, 0.02, 0.05]\), a moderate band \([0.1, 0.15, 0.2]\), or strong values \([0.3, 0.4, 0.5]\) to successive clients. For each schedule, the AT-coverage \(\gamma\) was swept through \(20\%\), \(50\%\), \(75\%\), and \(100\%\), after which robustness was probed with untargeted PGD attacks in the range \(\epsilon = 0.01\)–\(0.5\).

\begin{figure}
    \centering
    \includegraphics[width=\linewidth]{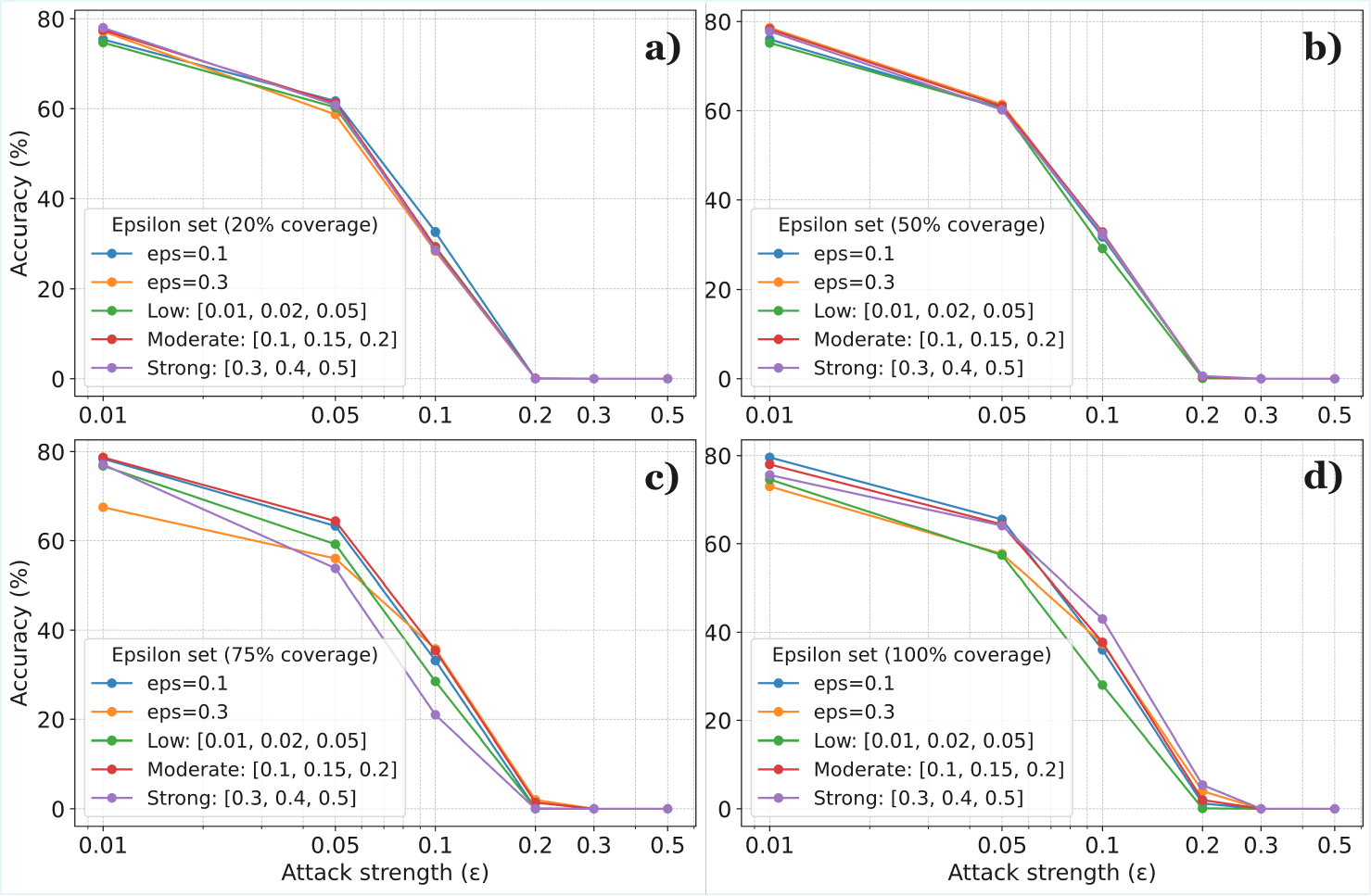}
    \caption{Robustness curves for the mixed‑$\epsilon$ adversarial‑training schedule. Lines denote coverage levels: (a) 20\% of clients adversarially trained, (b) 50\%, (c) 75\%, and (d) 100\%.}
    \label{fig:epsilon_set_coverage}
\end{figure}

Figure~\ref{fig:epsilon_set_coverage} illustrates the robustness curves across various coverage levels for each adversarial training method. At low coverage ($\le 50\%$), all methods achieve similar accuracy for light to moderate attacks ($0 \leq \epsilon \leq 0.1$) and collapse at $\epsilon \ge 0.2$, giving no meaningful robustness advantage. Hence, when only a minority of clients can be trained adversarially, the choice of $\epsilon$ approach has almost no impact on performance, as the model behaviour is driven by the majority of clean clients.

When coverage rises to $75\%$ (Fig.~\ref{fig:epsilon_set_coverage}c), meaningful gaps appear. Both the moderate mix and the $\epsilon = 0.1$ baseline hold strong light-attack accuracy ($\approx 64\%$ at $\epsilon = 0.05$) and retain $\approx 35\%$ at $\epsilon = 0.10$. In contrast, at $\epsilon = 0.10$, the strong mix collapses to $21\%$. Training on very large perturbations, therefore, erodes robustness to moderate attacks when coverage is high but not total.

At full coverage ($\gamma = 1$), all schedules benefit from universal AT, yet trade-offs shift (Fig.~\ref{fig:epsilon_set_coverage}d). It achieves the highest $\epsilon = 0.10$ accuracy ($43.0\%$), outstripping the moderate mix ($37.7\%$) and fixed $\epsilon = 0.1$ ($36.0\%$). Yet it also lowers the benign-input baseline ($\epsilon = 0$) to $77.3\%$, four points below $\epsilon = 0.1$. The moderate mix remains competitive, sitting midway in both the clean-like and moderate-attack regimes.

Aggregate metrics in Table~\ref{tab:robustness_ara_transposed_new} reinforce the curve-level story. The moderate mix records the highest Robustness Volume ($0.153$) and $\mathrm{ARA}_{\text{mean}}$ ($0.152$), thanks to its strong performance at $\gamma \ge 75\%$ and only a modest dip ($-0.005$) at $20\%$ coverage. The fixed $\epsilon = 0.1$ baseline delivers the most stable surface—second-best RV ($0.151$) and almost perfectly flat ARA across coverages—making it the safest single-$\epsilon$ choice. By contrast, the low-$\epsilon$ mix underperforms everywhere (RV $= 0.142$) and the high-$\epsilon$ baseline ($\epsilon = 0.3$) offers the lowest benign-input accuracy without compensating robustness gains.

When coverage is low or uncertain ($\gamma \le 50\%$), fine-tuning every adversarial client at a moderate fixed strength ($\epsilon \approx 0.1$) yields the most dependable accuracy-robustness trade-off. When a large majority of clients ($\gamma \approx 75\%$) can participate in AT, adopting a moderate mixed schedule achieves a balance between performance and robustness, providing clean-data accuracy comparable to $\epsilon = 0.1$ while maintaining enhanced resilience at $\epsilon \approx 0.10$. When full adversarial participation is used and the deployment environment is expected to face strong perturbations, the strong-$\epsilon$ mix provides the highest robustness, though it results in a modest reduction in clean-data accuracy.

\begin{table}[ht]
  \centering
  \caption{Robustness and ARA results for different epsilon splits}
  \label{tab:robustness_ara_transposed_new}
  \resizebox{\columnwidth}{!}{%
    \begin{tabular}{lccccc}
      \toprule
      Metric            & $\epsilon=0.1$ & $\epsilon=0.3$ & split [0.01,0.02,0.05] & split [0.1,0.15,0.2] & split [0.3,0.4,0.5] \\
      \midrule
      Robustness Volume & 0.151307       & 0.148319       & 0.142268               & 0.153087             & 0.145448            \\
      ARA$_{\text{mean}}$   & 0.150746       & 0.147972       & 0.141202               & 0.152338             & 0.147544            \\
      ARA$_{0\%}$       & 0.13634        & 0.13634        & 0.13634                & 0.13634              & 0.13634             \\
      ARA$_{20\%}$      & 0.15016        & 0.14227        & 0.14385                & 0.14605              & 0.14501             \\
      ARA$_{50\%}$      & 0.14949        & 0.15228        & 0.14400                & 0.15230              & 0.15137             \\
      ARA$_{75\%}$      & 0.15445        & 0.14895        & 0.14275                & 0.16146              & 0.12648             \\
      ARA$_{100\%}$     & 0.16329        & 0.16002        & 0.13907                & 0.16554              & 0.17852             \\
      \bottomrule
    \end{tabular}%
  }
\end{table}


\subsection{Non IID}

\begin{figure}
    \centering
    \includegraphics[width=\linewidth]{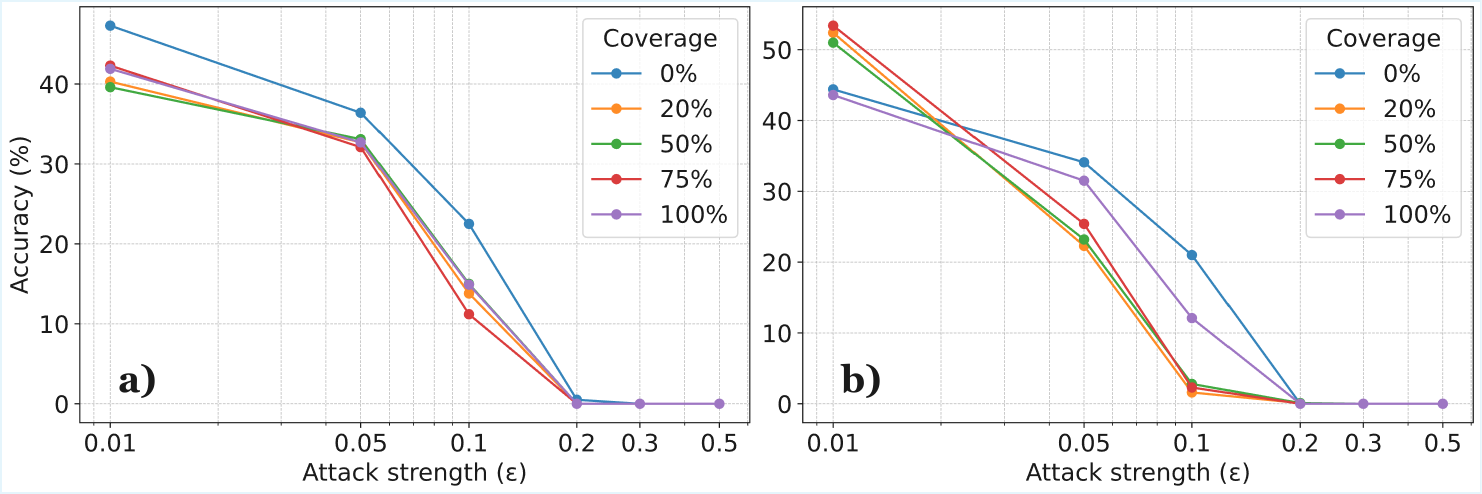}
    \caption{Model accuracy (\%) versus PGD attack strength $\epsilon$ for adversarially-trained networks with 0\%, 20\%, 50\%, 75\%, and 100\% coverage under sorted non-IID splits of (a) MNIST and (b) FMNIST.}
    \label{fig:Non_IID}
\end{figure}

As shown in Fig. \ref{fig:Non_IID}, our experiments with non‑IID MNIST and FMNIST datasets under sorted partitioning, for $0\%$ adversarial‑training coverage, model accuracy steadily fell as the perturbation strength ($\epsilon$) increased from low ($\varepsilon = 0.01$) to moderate ($\epsilon = 0.05$–$0.10$) and then to high ($\epsilon \ge 0.10$) levels. For coverages $\ge 20\%$, this drop was even more pronounced—especially under moderate attacks ($\epsilon = 0.05$–$0.10$), where accuracies fell to $1$–$12\%$ for FMNIST versus $21\%$ at $0\%$ coverage—showing a sharper loss of robustness. Across this range, all five adversarial‑training coverage curves stayed tightly grouped, with the $0\%$ line consistently performing best.
Under IID data, the same $\varepsilon = 0.10$ fine‑tune still achieves moderate accuracy and improves with more coverage. This confirms that the gains observed in IID settings disappear—and even reverse—when label‑homogeneous, non‑IID clients are introduced.

The numerical results in Table \ref{tab:mnist_fmnist_robustness_ara} match the visual trend in Fig. \ref{fig:Non_IID}. With $\epsilon = 0.10$, the Robustness Volume for MNIST falls from $0.151$ in the IID setting to $0.077$ in the sorted non‑IID setting, a $49\%$ reduction. Fashion‑MNIST shows a $60\%$ reduction ($0.154 \rightarrow 0.062$). The mean Adversarial Risk Area drops by almost the same proportion. Overall, moving from IID to non‑IID data removes about half of the model’s robustness surface.

The diminished effectiveness of adversarial training coverage in non-IID settings can be attributed to multiple underlying factors. First, sorted partitioning generates label-homogeneous clients, each handling a limited subset of labels. As a result, both the base model and any adversarial tweaks made to it become class-specific, and upon aggregation, the combined decision surface fails to generalize across the diverse range of perturbations encountered during training, thus hindering improvements in global robustness. 
Second, during the Federated Averaging (FedAvg) process, clients trained on disparate label subsets produce conflicting gradients. Increasing adversarial training coverage amplifies this conflict, leading to gradient cancellation rather than constructive model updates. Third, adversarial training introduces an additional regularization effect, which, when applied to data already challenged by distribution shifts, diverts the model from learning task-relevant features, consequently reducing baseline accuracy even at lower $\epsilon$ values such as 0.01.

Future work should address the compounded challenges of non-IID data and adversarial robustness in quantum-federated learning, where robustness volume halves compared to classical IID settings. A promising direction is exploring client-shared reference sets—even small, synthetic, or IID-like subsets—to stabilize aggregated models and mitigate adversarial vulnerabilities under extreme data heterogeneity. Additionally, the breakdown of robustness transfer at high noise levels ($\epsilon \ge 0.20$) suggests revisiting adaptive training strategies, such as curriculum learning or mixed-$\epsilon$ schedules, to replace rigid fixed-$\epsilon$ frameworks. These approaches could dynamically adjust perturbation magnitudes during training, broadening robustness generalization under strong non-IID conditions while balancing security and model performance.

\begin{table}[ht]
  \centering
  \caption{Robustness Volume and ARA mean under IID vs.\ non‐IID settings}
  \label{tab:mnist_fmnist_robustness_ara}
  \begin{tabular}{lcc}
    \toprule
    Experiment                      & Robustness Volume & ARA\_mean       \\
    \midrule
    IID MNIST                       & 0.151             & 0.151           \\
    Non‐IID MNIST (sorted)          & 0.077 (–49\%)     & 0.079 (–48\%)   \\
    IID FMNIST                      & 0.154             & 0.155           \\
    Non‐IID FMNIST (sorted)         & 0.062 (–60\%)     & 0.066 (–57\%)   \\
    \bottomrule
  \end{tabular}
\end{table}

\section{Conclusion}

Based on the experimental analysis, this work shows that robust quantum federated learning depends on balancing adversarial training coverage, perturbation strength, and client participation. Partial adversarial coverage (20\%--50\%) improves resilience against moderate attacks ($\epsilon \leq 0.1$) without harming clean accuracy, proving that full participation isn't always needed. Fine-tuning is more effective than training from scratch, especially under moderate threats and partial coverage, as it keeps accuracy stable. Mixed-$\epsilon$ schedules (like $[0.1, 0.15, 0.2]$) work best when client coverage is high ($\geq 75\%$), while a fixed moderate $\epsilon$ ($\sim 0.1$) is better for low coverage. However, non-IID data remains a key challenge, reducing robustness significantly and sometimes even harming accuracy, showing that data distribution is critical in QFL design.

Therefore, effective QFL depends on aligning clients and adversaries by selecting appropriate coverage levels and perturbation schedules that match the expected threat models and data characteristics. Looking ahead, future efforts must focus on client-aware defenses, such as personalized $\epsilon$ schedules or shared reference sets, to reduce vulnerabilities made worse by non-IID data while managing the trade-off between quantum robustness and accuracy.

\section*{Acknowledgments}
This work was supported in part by the Center for Cyber Security (CCS), funded by Tamkeen under the NYUAD Research Institute Award G1104.

\bibliographystyle{IEEEtran}
\bibliography{refs_}
\end{document}